\documentclass[aps,prl,twocolumn,groupedaddress, showpacs, floatfix]{ revtex4}
\usepackage{amssymb}
\usepackage{amsmath}
\usepackage{graphicx}
\begin{document}

\title{ Structure of smectic defect cores: an
X-ray study of 8CB liquid crystal ultra-thin films}
\author{Jean-Philippe MICHEL}
\author{Emmanuelle LACAZE}
\email[]{emmanuelle.lacaze@insp.jussieu.fr}
\author{Michel GOLDMANN}
\affiliation{INSP, Universit\'es Paris 7 et 6, UMR-CNRS 7588, Campus Boucicaut,
140 rue de Lourmel, F-75015 Paris, FRANCE}
\author{Marc GAILHANOU}
\affiliation{LURE,  Bat 209D, UniversitŽ Paris Sud, F-91405  Orsay CEDEX,
FRANCE}
\author{Marc de BOISSIEU}
\affiliation{ LTPCM,  INPG, BP 75, 38402 Saint Martin d'H\`eres, FRANCE}
\author{Michel ALBA}
\affiliation{ LLB, UMR12 CEA-CNRS, CEA-Saclay,
F-91191 Gif-sur-Yvette Cedex, FRANCE}

\date{\today}

\begin{abstract}

We study the structure of  very thin liquid crystal films frustrated by antagonistic
anchorings in the smectic phase.
In a cylindrical geometry, the structure is dominated by the defects for
film thicknesses smaller than 150 nm
and the detailed topology of the defects cores can be revealed  by x-ray diffraction.
They appear to be split in half
tube-shaped Rotating Grain Boundaries (RGB). 
We determine the RGB spatial extension  and  
evaluate its energy per unit line.
Both are significantly larger than the ones usually proposed in the literature.
\end{abstract}

\pacs{ 61.10.-i,68.35.Bs, 61.30, 68.35.Md}

\maketitle


The combined technological
interest for liquid crystal (LC) devices and for small size devices
requires now a precise understanding of the microscopic structure of
liquid crystalline films. Progress in the LC physics is often intimately connected to 
the understanding of the LC defects \cite{Frie22}. If now the LC 
defects are precisely described from a macroscopic point of view, 
a microscopic description is still lacking, with only simulations
\cite{Mka00} currently available. The use of powerful 
techniques such as x-ray diffraction performed with synchrotron 
radiation sources allows now to bridge such a gap.
In lamellar phases (e.g. smectic phases in thermotropic systems),
the most common defects are focal conics. They are singularity lines
where the layer curvature is not defined, around which the layers rotate.
They are ellipses (degenerated into 
straight lines, the disclinations, in "oily streaks") conjugated with 
hyperbolae (degenerated into curvature
walls, in "oily streaks") \cite{Frie22, Kle77, Kle04}.
The inner structure of these singularity lines, 
remains unknown since they are buried within the 
deformed film and only a very small amount of matter is involved. 
To overcome this first problem, we have studied very thin smectic films 
in which linear  defects dominate the film structure.
In this limit, the defects can no longer be considered
as singularity lines but are split into two-dimensional structures.
We have used oriented samples of "oily streaks", self-assembled in regular stripes,
and we have performed  x-ray diffraction experiments at synchrotron sources
that allowed us to determine the rotation of the layers close to the defects.
We have revealed the topological structure of the disclination core. The spatial 
extension has been determined and the energy per unit of 
line estimated. Both values are at least one order of magnitude larger  than
the usual estimations proposed in the literature.  
\begin{figure}[!ht]
\includegraphics[width=0.85\columnwidth]{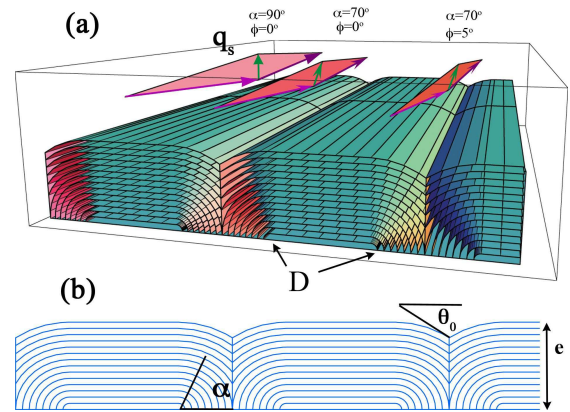}
\caption{\label{fig1} (a) Scheme of layers,
concentrically stacked into flattened
hemicylinders lying flat on the substrate, with half disclinations
located at points D.
Examples of diffusion
triangles for various $\alpha$ and $\phi$ angles are drawn, 
with enhanced incident angles for clarity, and momentum transfer, $\vec{q}_{s}$,
in green. (b) Flattened
hemicylinders, presented in the plane perpendicular
to the hemicylinder axes:
The curvature walls between
neighboring hemicylinders
are characterized by the angle at the top, denoted by $\theta_{0}$.
The curvature of the layers is associated with the $\alpha$ angle.}
\end{figure}

We have studied 
8CB (4-n-octyl-4'-cyanobiphenyl) smectic films adsorbed on a MoS$_{2}$ substrate, 
deformed through strong
antagonistic anchorings at both interfaces but ordered by the single crystal 
surface of MoS$_{2}$ (Molybdenum disulfide).
The relaxation of the constraints imposed by the antagonistic anchorings
occurs through the formation of a periodic network of flattened
hemicylinders, parallel to the substrate  \cite{Mic04}. These hemicylinders are associated with 
half-disclinations locked on the substrate
at the centers of curvature of the quarters of cylinders (points D in figure \ref{fig1}),
conjugated with curvature walls.

The  8CB, smectic in bulk at 25$^{o}$C, is used
without any further purification (BDH-Germany).
A 0.1 mol/l solution of 8CB in dichloromethane is deposited  on a freshly
cleaved surface of MoS$_{2}$. The film's thickness, $e$, is controlled
by spin coating at a speed varying between 1000 and 6000
rpm.
The film thickness is checked by optical microscopy and 
determined by the film color, according to the Newton tint table,
with an average error of 15 nm in the 70-350 nm range.
The sample is annealed at $80^{o}C$
to allow the formation of an ordered 8CB/MoS$_{2}$ interface
which imposes a strong planar
unidirectional anchoring within large domains \cite{Laca04}, antagonistic to the
homeotropic anchoring at the
8CB/air interface. 
\begin{figure}[!ht]
\includegraphics[width=0.7\columnwidth]{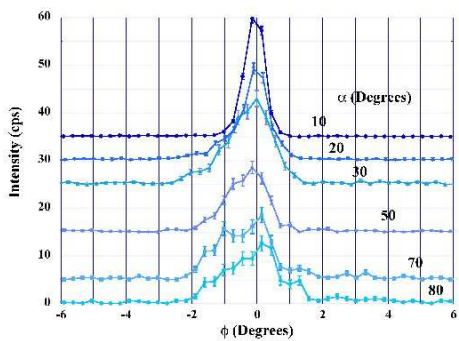}
\caption{\label{fig2} Bragg intensity  variation with $\phi$
for different $\alpha$ values 
(e = 200 nm). The narrowing at small $\alpha$ values
reveals a weaker mosaicity and is related to the strong anchoring on the substrate.}
\end{figure}

X-ray diffraction experiments are performed on the  D2AM
(ESRF, Grenoble, France) and H10 (LURE, Orsay, France) synchrotron beamlines.
The energy is fixed at 8 keV, the beam spot is 50x 50 $\mu m^{2}$
large.  We take
advantage of the periodic character of the smectic A phase and detect
the first Bragg peak associated with the 8CB
bulk period ($q_{s}=0.198 \AA^{-1}$). In our resolution limited set-up,
after geometrical corrections (the background is subtracted,
the evolution of the beam footprint and of the penetration depth
leading to a transmission factor step are taken into account \cite{Mic04} ),
the Bragg intensity becomes proportional to
the number of layers oriented with the director
parallel to the wave-vector, $\vec{q}_{s}$.
The orientation of the layers
is then followed by tilting $\vec{q}_{s}$ and measuring
the evolution of the Bragg intensity. The $\phi$ angle is defined
as the angle between the diffraction plane and the hemicylinder axes.
The $\alpha$ angle is defined in the plane perpendicular to
the hemicylinder axes as the disorientation of $\vec{q}_{s}$ and
such as the disorientation of the layers,
with respect to the substrate surface (fig \ref{fig1}b).
We have first checked that the cylindrical symmetry imposed by the
anchoring antagonism is preserved within the smectic film
whatever the $e$ values: The $\phi$ scans of the Bragg intensity
remain centered at $\phi = 0$, whatever the $\alpha$ values
(see fig. \ref{fig2} for e = 200 nm). We have then measured the
Bragg intensity as a function of
$\alpha$ for $\phi = 0$,
$\alpha=0^{o}$ corresponding to the layers perpendicular
to the substrate and $\alpha=90^{o}$ corresponding to the
layers parallel to it.
\begin{figure}[!ht]
\includegraphics[width=0.85\columnwidth]{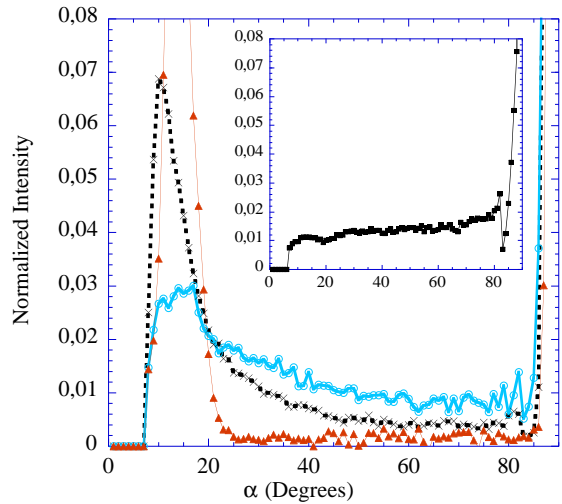}
\caption{\label{fig3} Bragg intensity  variation with $\alpha$  after geometrical correction
and background  subtraction.
Intensity is normalized
with respect to the integrated intensity from $ 0^{o} \leq \alpha  \leq 85^{o}$.
Curves for 
thin films  are shown in the main frame (e = 70, 150 and 200 nm [full triangles,
crosses and open circles]), and for a thicker film 
in the insert (e=450 nm [full squares]).}
\end{figure}
The figure \ref{fig3} presents the evolution when $e$
decreases from 450 nm to 70 nm.
All curves are  normalized
with respect to the integrated intensity for $0^{o} \leq \alpha  \leq 85^{o}$.
Thus the main
frame presents the distribution of rotating
layers between $\alpha=0^{o}$ and $\alpha=85^{o}$, for
e = 200, 150 and 70 nm. The null intensity from $\alpha=0^{o}$ to
$\alpha=7^{o}$ is associated with the
8CB critical angle ($\theta_{8CB} = 0.17^{o}$). When $\alpha$ becomes smaller
than 7$^{o}$,
the incident beam tilt angle becomes smaller than $\theta_{8CB}$ and the
beam penetration
becomes negligible.
The rapid increase at $85^{o}$  corresponds to the layers parallel
to the  substrate and is present for all e. 
In the inset, the quasi-constant Bragg intensity, for e = 450 nm,
illustrates that, in case of thick film, only the signal of the bulk
organized in flattened hemicylinders is detected, associated with a
quasi-constant distribution of rotating layers from $\alpha=0^{o}$ to
$\alpha=85^{o}$ \cite{Mic04}.
The pronounced evolution of the distribution shape as $e$ decreases, discloses 
the increasing influence of the half disclination inner structure:
The number of rotating layers 
appears larger at small $\alpha$ than at large $\alpha$.
For very small thicknesses (e = 70 nm), 
the x-ray signal is controlled by the half disclination structure,
leading to
Bragg intensity  only detected for
$7^{o} \leq \alpha \leq 25^{o}$ and at $\alpha=90^{o}$.
If $e \geq 150 nm$, layers rotating up to
$\alpha=85^{o}$ are measured, indicating that the half disclination size
is smaller than 150 nm.
The negligible half disclination contribution to
the x-ray data, if e = 450 nm, indicates that its size does not
vary significantly with the thickness.

\begin{figure}
\includegraphics[width=0.9\columnwidth]{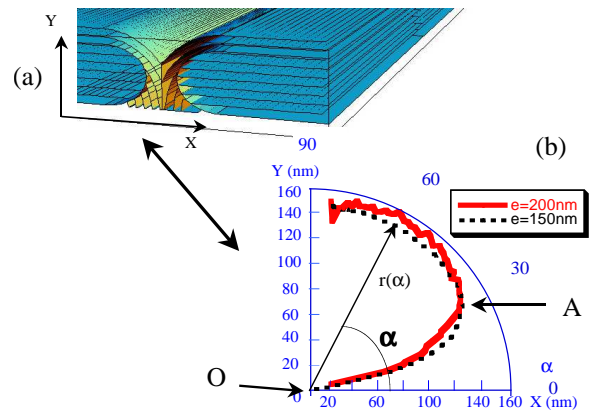}
\caption{\label{fig4}(a) Scheme of enlarged flattened hemicylinders
close to the half disclinations. The
arrow associates a schematized half disclination split in RGB with the profile extracted from the
x-ray data:
(b) RGB profile, as deduced
from the x-ray data in cases of
thicknesses 150 nm (dotted-line) and 200 nm (line)
and a value $\theta_{0}=60^{o}$,  presented in the plane perpendicular
to the hemicylinder axes.}
\end{figure}

The gap observed in 70 nm thick films,
between flat layers ($\alpha=90^{o}$) and
rotating ones ($7^{o} \leq \alpha \leq 25^{o}$),
associated with the still preserved cylindrical symmetry of the film,
allows to
model the split mode of the half disclination by a Rotating Grain Boundary (RGB).
This RGB
partitions the parallel layers from the rotating ones (figure \ref{fig4}a)
and indeed eliminates a larger number of rotating layers at high $\alpha$ values than
at low $\alpha$ ones.

In order to extract the RGB shape from the x-ray data, 
we define $r(\alpha)$ the polar coordinate
of the RGB, taken from the center of curvature, O, of the quarter of 
cylinder (fig. \ref{fig4}b).
Taking in account the
curvature walls between the quarters of cylinders,
characterized by the upper $\theta_{0}$ angle, defined in fig. \ref{fig1}b,
the number of layers $n(\alpha )$
rotating from $\alpha$ to $\alpha+1^{o}$ is evaluated:

\begin{equation}
\begin{array}{lll}
n(\alpha )&\propto \frac{e \sin \theta_{0}}{\cos \alpha}-{ r(\alpha)}
 &  \text{for } \alpha \leq 90^{o}-\theta_{0} \\
 & \propto{ e} - r(\alpha)
  &  \text{for }  \alpha \geq 90^{o}-\theta_{0}
\end{array}
\end{equation}

The number of layers is also proportional to $i(\alpha )$, the x-ray 
intensity. The observed high intensity
at small $\alpha$ in fig. \ref{fig3} shows that the RGB spatial extension in the
direction associated with $\alpha$ close to 0$^{o}$ is negligible and the
proportionality between $n(\alpha )$ and $i(\alpha )$ is obtained,
using a small $\alpha_0$ value, however higher than 7$^{o}$:
$n(\alpha_0)\propto (e \frac{\sin \theta_{0}}{\cos \alpha_0}) \approx A i(\alpha_0)$,
leading to

\begin{equation}
\begin{array}{lll}
r(\alpha) & = e \frac { \sin (\theta_{0})}{\cos( \alpha_{0})}
( \frac {\cos( \alpha_{0})} {\cos(\alpha)}
-    \frac { i(\alpha)} {i(\alpha_0)}
)
 & \alpha \leq 90^{o}-\theta_{0} \\
 
 & = e  (1 -\frac{ i(\alpha) \sin (\theta_{0}) } 
{i(\alpha_0)\cos (\alpha_0)} )
 & \alpha \geq 90^{o}-\theta_{0}
\end{array}
\end{equation}

We have to estimate now the characteristic $\theta_{0}$ angle.
The observed quasi-constant
evolution of x-ray data between $\alpha=50^{o}$ and $\alpha=85^{o}$
suggests that $\theta_{0} \geq 40^{o}$ for $200 nm \leq e \leq 150 nm$. 
Postulating a value
$\theta_{0}=60^{o}$, two similar profiles of the RGB 
are obtained (figure \ref{fig4}b) for e = 200 nm and e = 150 nm, as expected.
These profiles have been extrapolated continuously between $\alpha=7^{o}$
and $\alpha=0^{o}$ down to the O  point.
They are only slightly modified, when varying the $\theta_{0}$ value between $\theta_{0}=90^{o}$
and $\theta_{0}=40^{o}$.

The contribution of this RGB profile to the x-ray intensity can now be 
calculated, using the reverse procedure. It is found negligible for e = 450 nm,
in agreement with the
experimental results. On the contrary, the evolution towards ultra-thin
films is driven by the RGB
presence. The geometry of films thinner than 80 nm can be
deduced from the profile of figure \ref{fig4}b. They
are composed of flat layers alternating with quasi-perpendicular ones,
in agreement with the
x-ray results on the
70 nm thick film.
It should be noted that these high disorientations are
present whatever the film thickness, since they always exist at
the bottom of the buried defects.

The RGB can be described as an half-tube with a
quasi-elliptical cut of small axis equal to 140 nm and of half-large axis
equal to 110 nm (deduced from fig. \ref{fig4}b). Such a spatial
extension is significantly
larger than  $\lambda$,
the smectic penetration length,
equal to some nanometers, the distance usually associated
with the focal conic cores in the literature. The
origin of focal conic cores is related to
the replacement of elastic curvature by dilation. A 
split mode in grain boundaries, characterized by local dilation of
the layers, should then be a general property of focal conics in general. Such a 
replacement should be independent of the film thickness,
as experimentally observed, as well as it should be independent
of the anchoring energy values at the interfaces.
It should mainly depend on the ratio between the elastic and 
compression modulus and on the disorientations imposed by the 
external constraints \cite{Wil75}, leading to large spatial extensions,
similarly for focal conics of different systems.

However one expects the width of the RGB to be of the order of
$\lambda$. An exact value of the width can not be
obtained through x-ray measurements and depends on the detailed
structure of the RGB. The RGB is definitely not a simple grain boundary because it
can not account for the topological continuity of the layers,
contrary to straight curvature walls 
\cite{Wil75, Blan99}.
Indeed, at point A of the RGB (see figure \ref{fig4}b),
20 flat layers have to
accommodate 45 rotating layers. This necessitates 20 dislocations
between the start of the RGB (point O) and point A,
as well as between point A and the
other RGB extremity. The simplest models for the RGB correspond, either to
a grain boundary with many dislocations
(even if they are not detected though x-ray diffraction or optical microscopy
measurements \cite{Amb02}),
or to a nematic envelope along the RGB \cite{Doz}.
Following these two assumptions, the energy per
unit line of the RGB can be calculated and associated to the
half-disclination core energy.

In the dislocation hypothesis,
this energy is calculated using the energy of a simple grain boundary
\cite{Blan99}, increased by 40 times the dislocation energy term, of 
the order of $K_{8CB}$
\cite{Kle77} ($K_{8CB}= 11\times 10^{-12} J m^{-1}$ being the smectic
8CB curvature modulus \cite{Brad85}):

\begin{equation}
\begin{array}{lll}
E_{RGB} &= & 40 K_{8CB} + 1/2 \int_{0}^{\frac{\pi}{2}} \frac{K}{\lambda}
r(\alpha) \frac{\cos(\frac{\pi}{2} -\alpha)}{2}\\
 & &(\tan(\pi/2 -\alpha)/2-(\pi/2 -\alpha)/2) d\alpha \\
 &=  & 42 K_{8CB}
\end{array}
\end{equation}
leading to an energy per unit of line
mainly imposed by the dislocations energy. 

In the hypothesis of a nematic envelope,
the RGB energy can also be estimated,
considering that the nematic envelope replaces a smectic envelope
whose order parameter is equal to the smectic bulk order
parameter:
\begin{equation}
\begin{array}{lll}
E_{RGB} &= P_{RGB}({a\over 2}\Psi^{2}(T_{0}-T)^{s}-{b\over4}\Psi^{4}) \delta \\
       &=P_{RGB}  (a^{2}(T-T_{0})^{2s})/4b) \delta  \\
       & =108 K_{8CB}
\end{array}
\end{equation}
$P_{RGB} = 350 nm$ is the RGB perimeter, $\delta$ is 
the thickness envelope taken as equal to 1 nm, $\Psi$ is the smectic
order parameter, $T_{0}$ the
transition temperature ($T_{0} = 33.5^{o}$ C), a and b
the second- and fourth-order coefficients, s the critical exponent
in the de-Gennes free energy
versus $\Psi$ and temperature
($a^{2}$/b = 3.3 $10^{6} J K^{-0.68} m^{-3}$, $s=0.34$ \cite{Tho82}).

These two simple evaluations of the RGB energy  are
of the same order of magnitude  (42-108*K).
Improved approximations would call for a
finer description of the RGB structure. In the first model,
the dislocation distribution along the RGB should be taken
into account and, in the second model, a continuous evolution of the
smectic order parameter \cite{Kra93}.
These calculations demonstrate that the energy per unit of line
is at least one order of
magnitude higher
than K, the elastic modulus, which is the value usually proposed in the literature
for a focal conic core energy per unit of line \cite{Gen,
Li94, Kle00}. On the contrary, this result is consistent with a value of 15*K \cite{Blan99},
indirectly obtained from
the observations of toroidal focal conics in a lamellar/sponge system.
Both results suggest then that
disclinations and focal conic core energies
per unit of line are closer to 10-100*K than to K, the exact value
depending on the geometries of the different systems.

X-ray measurements of thin smectic films deformed through a $90^{o}$
antagonism of anchoring at both interfaces have revealed the
structure of half disclination cores. They
appear to be split in half-tubed Rotating Grain Boundaries (RGB),
with a  spatial extension
of 140x110 nm$^{2}$
and an energy per unit of line between 40*K and 110*K, K being the elastic modulus.
Both values are more than one order of magnitude higher than the values
usually proposed
in the literature. As for focal conics, the half-disclination structure
is imposed by the replacement
of elastic curvature by dilation.
In case of similar ratios between compression and elastic modulus,
different systems submitted to high disorientations
should also present split modes of the cores in  grain boundaries
of large spatial extensions. 
The high energy per unit of line
is due to the topological discontinuity of the smectic layers
from either part of the RGB. This discontinuity can be accounted either by a
grain boundary with a high density of dislocations, or
by the presence of a nematic envelope. Such high energies per unit of line
could constitute a general feature of
focal conics and disclinations,
whereas only straight grain boundaries allow
topological connections between the layers.
Our results not only can be generalized to different defects of 
lamellar phases (for example, one can think of copolymer or amphiphilic systems
that are of fundamental and technological importance in many applications)
but also demonstrate the ability of x-ray 
diffraction to probe new soft matter structures. It is now possible,
using synchrotron radiation 
sources, to reveal the detailed inner structure of defects, buried into 
smectic films.

\begin{acknowledgments}
We thank  Maurice Kl\'eman and Bernard Croset
for useful discussions.
\end{acknowledgments}


\begin{thebibliography}{19}
\expandafter\ifx\csname natexlab\endcsname\relax\def\natexlab#1{#1}\fi
\expandafter\ifx\csname bibnamefont\endcsname\relax
  \def\bibnamefont#1{#1}\fi
\expandafter\ifx\csname bibfnamefont\endcsname\relax
  \def\bibfnamefont#1{#1}\fi
\expandafter\ifx\csname citenamefont\endcsname\relax
  \def\citenamefont#1{#1}\fi
\expandafter\ifx\csname url\endcsname\relax
  \def\url#1{\texttt{#1}}\fi
\expandafter\ifx\csname urlprefix\endcsname\relax\def\urlprefix{URL }\fi
\providecommand{\bibinfo}[2]{#2}
\providecommand{\eprint}[2][]{\url{#2}}

\bibitem[{\citenamefont{Friedel}(1922)}]{Frie22}
\bibinfo{author}{\bibfnamefont{G.}~\bibnamefont{Friedel}},
  \bibinfo{journal}{Ann. Phys. (Paris)} \textbf{\bibinfo{volume}{18}},
  \bibinfo{pages}{273} (\bibinfo{year}{1922}).
  
    \bibitem[{\citenamefont{Mkaddem and Gartland}(2000)}]{Mka00}
\bibinfo{author}{\bibfnamefont{S.} \bibnamefont{Mkaddem}} \bibnamefont{and}
  \bibinfo{author}{\bibfnamefont{E.~C.}~\bibnamefont{Gartland}},
  \bibinfo{journal}{Phys. Rev. E} \textbf{\bibinfo{volume}{62}},
  \bibinfo{pages}{6694} (\bibinfo{year}{2000}).
  
\bibitem[{\citenamefont{Kl\'eman}(1977)}]{Kle77}
\bibinfo{author}{\bibfnamefont{M.}~\bibnamefont{Kl\'eman}},
  \emph{\bibinfo{title}{Introduction to Liquid Crystals}}
  (\bibinfo{publisher}{Les editions de physique}, \bibinfo{address}{Paris},
  \bibinfo{year}{1977}).
  
  \bibitem[{\citenamefont{Kl\'eman, Lavrentovich and Nastishin}(2004)}]{Kle04}
\bibinfo{author}{\bibfnamefont{M.}~\bibnamefont{Kl\'eman}},
  \bibinfo{author}{\bibfnamefont{O.~D.} \bibnamefont{Lavrentovich}} \bibnamefont{and}
  \bibinfo{author}{\bibfnamefont{Yu. A.}~\bibnamefont{Nastishin}}
  \emph{\bibinfo{title}{Dislocation and disclination in mesomorphic phases}}
  (\bibinfo{publisher}{F.R.N. Nabarro and J.P. Hirth}, \bibinfo{address}{Paris},
  \bibinfo{year}{2004}).

\bibitem[{\citenamefont{Michel et~al.}(2004)\citenamefont{Michel, Lacaze, Alba,
  de~Boissieu, Gailhanou, and Goldmann}}]{Mic04}
\bibinfo{author}{\bibfnamefont{J.~P.} \bibnamefont{Michel}},
  \bibinfo{author}{\bibfnamefont{E.}~\bibnamefont{Lacaze}},
  \bibinfo{author}{\bibfnamefont{M.}~\bibnamefont{Alba}},
  \bibinfo{author}{\bibfnamefont{M.}~\bibnamefont{de~Boissieu}},
  \bibinfo{author}{\bibfnamefont{M.}~\bibnamefont{Gailhanou}},
  \bibnamefont{and} \bibinfo{author}{\bibfnamefont{M.}~\bibnamefont{Goldmann}},
  \bibinfo{journal}{Phys. Rev. E} \textbf{\bibinfo{volume}{70}},
  \bibinfo{pages}{011709} (\bibinfo{year}{2004}).

\bibitem[{\citenamefont{Lacaze et~al.}(2004{\natexlab{b}})\citenamefont{Lacaze,
  Michel, Goldmann, de~Boissieu, Gailhanou, and Alba}}]{Laca04}
\bibinfo{author}{\bibfnamefont{E.}~\bibnamefont{Lacaze}},
  \bibinfo{author}{\bibfnamefont{J.~P.} \bibnamefont{Michel}},
  \bibinfo{author}{\bibfnamefont{M.}~\bibnamefont{Goldmann}},
  \bibinfo{author}{\bibfnamefont{M.}~\bibnamefont{de~Boissieu}},
  \bibinfo{author}{\bibfnamefont{M.}~\bibnamefont{Gailhanou}},
  \bibnamefont{and} \bibinfo{author}{\bibfnamefont{M.}~\bibnamefont{Alba}},
  \bibinfo{journal}{Phys. Rev. E} \textbf{\bibinfo{volume}{69}},
  \bibinfo{pages}{041705} (\bibinfo{year}{2004}{\natexlab{b}}).

  
\bibitem[{\citenamefont{Williams and Kl\'eman}(1975)}]{Wil75}
\bibinfo{author}{\bibfnamefont{C.~E.} \bibnamefont{Williams}} \bibnamefont{and}
  \bibinfo{author}{\bibfnamefont{M.}~\bibnamefont{Kl\'eman}},
  \bibinfo{journal}{J. Phys. C1-sup3} \textbf{\bibinfo{volume}{36}},
  \bibinfo{pages}{315} (\bibinfo{year}{1975}).

\bibitem[{\citenamefont{Blanc and Kl\'eman}(1999)}]{Blan99}
\bibinfo{author}{\bibfnamefont{C.}~\bibnamefont{Blanc}} \bibnamefont{and}
  \bibinfo{author}{\bibfnamefont{M.}~\bibnamefont{Kl\'eman}},
  \bibinfo{journal}{Eur. Phys. J. B} \textbf{\bibinfo{volume}{10}},
  \bibinfo{pages}{53} (\bibinfo{year}{1999}).

\bibitem[{\citenamefont{Ambrozic et~al.}(2002)\citenamefont{Ambrozic, Kralj,
  and Zumer}}]{Amb02}
\bibinfo{author}{\bibfnamefont{M.}~\bibnamefont{Ambrozic}},
  \bibinfo{author}{\bibfnamefont{S.}~\bibnamefont{Kralj}}, \bibnamefont{and}
  \bibinfo{author}{\bibfnamefont{S.}~\bibnamefont{Zumer}},
  \bibinfo{journal}{Eur. Phys. J. E} \textbf{\bibinfo{volume}{8}},
  \bibinfo{pages}{413} (\bibinfo{year}{2002}).

\bibitem[{\citenamefont{Dozov and Durand}(1994)}]{Doz}
\bibinfo{author}{\bibfnamefont{I.}~\bibnamefont{Dozov}} \bibnamefont{and}
  \bibinfo{author}{\bibfnamefont{G.}~\bibnamefont{Durand}},
  \bibinfo{journal}{Europhys. Lett.} \textbf{\bibinfo{volume}{28}},
  \bibinfo{pages}{25} (\bibinfo{year}{1994}).


\bibitem[{\citenamefont{Bradshaw and Raynes}(1985)}]{Brad85}
\bibinfo{author}{\bibfnamefont{M.~J.} \bibnamefont{Bradshaw}} \bibnamefont{and}
  \bibinfo{author}{\bibfnamefont{E.~P.} \bibnamefont{Raynes}},
  \bibinfo{journal}{J. Physique} \textbf{\bibinfo{volume}{46}},
  \bibinfo{pages}{1513} (\bibinfo{year}{1985}).

\bibitem[{\citenamefont{Thoen et~al.}(1985)\citenamefont{Thoen, Marynissen, and
  Van~Dael}}]{Tho82}
\bibinfo{author}{\bibfnamefont{J.}~\bibnamefont{Thoen}},
  \bibinfo{author}{\bibfnamefont{H.}~\bibnamefont{Marynissen}},
  \bibnamefont{and} \bibinfo{author}{\bibfnamefont{W.}~\bibnamefont{Van~Dael}},
  \bibinfo{journal}{Phys. Rev. A} \textbf{\bibinfo{volume}{46}},
  \bibinfo{pages}{1513} (\bibinfo{year}{1985}).

\bibitem [{\citenamefont{Kralj et~al.}(1993)\citenamefont{Kralj and Sluckin}}]{Kra93}
\bibinfo{author}{\bibfnamefont{S.}~\bibnamefont{Kralj}}
  \bibnamefont{and} \bibinfo{author}{\bibfnamefont{T.~J.}~\bibnamefont{Sluckin}},
  \bibinfo{journal}{Phys. Rev. E} \textbf{\bibinfo{volume}{48}},
  \bibinfo{pages}{R3244} (\bibinfo{year}{1993}).
  
\bibitem[{\citenamefont{de~Gennes}(1972)}]{Gen}
\bibinfo{author}{\bibfnamefont{P.~G.} \bibnamefont{de~Gennes}},
  \bibinfo{journal}{C. R. Acad. Sc. Paris} \textbf{\bibinfo{volume}{275}},
  \bibinfo{pages}{549} (\bibinfo{year}{1972}).

\bibitem[{\citenamefont{Li and Lavrentovich}(1994)}]{Li94}
\bibinfo{author}{\bibfnamefont{Z.}~\bibnamefont{Li}} \bibnamefont{and}
  \bibinfo{author}{\bibfnamefont{O.~D.} \bibnamefont{Lavrentovich}},
  \bibinfo{journal}{Phys. Rev. Lett.} \textbf{\bibinfo{volume}{73}},
  \bibinfo{pages}{280} (\bibinfo{year}{1994}).

\bibitem[{\citenamefont{Kl\'eman and Lavrentovich}(2000)}]{Kle00}
\bibinfo{author}{\bibfnamefont{M.}~\bibnamefont{Kl\'eman}} \bibnamefont{and}
  \bibinfo{author}{\bibfnamefont{O.~D.} \bibnamefont{Lavrentovich}},
  \bibinfo{journal}{Eur. Phys. J. E} \textbf{\bibinfo{volume}{2}},
  \bibinfo{pages}{47} (\bibinfo{year}{2000}).

\end{thebibliography}
\end{document}